# BAyesian Bent-Line Regression model for longitudinal data with an application to the study of cognitive performance trajectories in Wisconsin Registry for Alzheimer's Prevention

Lianlian Du[a,c], Rebecca Langhough Koscik[a,b,e], Tobey J Betthauser[a,b,e], Sterling C. Johnson[a,b,e,f], Bret Larget[d] and Rick Chappell[b,c,d]*

[a]*Wisconsin Alzheimer's Institute, University of Wisconsin-Madison School of Medicine and Public Health, Madison, WI, USA;* [b]*Wisconsin Alzheimer's Disease Research Center, Madison, WI, USA;* [c]*Department of Biostatistics and Medical Informatics, School of Medicine and Public Health, University of Wisconsin-Madison, Madison, WI, USA;* [d]*Department of Statistics, University of Wisconsin-Madison, Madison, WI, USA;* [e]*Department of Medicine, University of Wisconsin-Madison School of Medicine and Public Health, Madison, WI, USA;* [f]*Madison VA GRECC, William S. Middleton Memorial Hospital, Madison, WI, USA*

*corresponding author:

chappell@biostat.wisc.edu

# BAyesian Bent-Line Regression model for longitudinal data with an application to the study of cognitive performance trajectories in Wisconsin Registry for Alzheimer's Prevention

**Abstract:** Preclinical Alzheimer's disease (AD), the earliest stage in the AD continuum, can last fifteen to twenty years, with cognitive decline trajectories nonlinear and heterogeneous between subjects. Characterizing cognitive decline in the preclinical phase of AD is critical for the development of early intervention strategies when disease-modifying therapies may be most effective. In the last decade, there has been an increased interest in the application of change point (CP) models to longitudinal cognitive outcomes. Because patients' change points can vary greatly, it is essential to model this variation. In this paper, we introduce a BAyesian Bent-Line Regression model longitudinal data on cognitive function in middle-aged adults with a high risk of AD. We provide an approach for estimating the fixed (group-level) and random (person-level) CPs, slopes pre- and post-CP, and intercepts at CP for cognition. Our model not only estimates the individual cognitive trajectories but also the distributions of the cognitive bent line curves at each age, enabling researchers and clinicians to estimate subjects' quantiles. Simulation studies show that the estimation and inferential procedures perform reasonably well in finite samples. The practical use is illustrated by an application to a longitudinal cognitive composite in the Wisconsin Registry for Alzheimer's Prevention (WRAP).

# 1 Introduction

Alzheimer's disease (AD), the leading cause of dementia globally (Livingston et al., 2017), is a progressive, neurodegenerative disease associated with cognitive, functional, and behavioural impairments. AD follows a progressive disease continuum that extends from an asymptomatic phase with biomarker evidence of AD pathological changes (preclinical AD), through minor cognitive (mild cognitive impairment [MCI]) and/or neurobehavioral (mild behavioural impairment [MBI]) changes to, ultimately, AD dementia (Porsteinsson et al., 2021). NIA-AA Research Framework defined the staging systems to categorize AD across this continuum (Jack et al., 2018). Studies with potentially disease-modifying drugs conducted at the mild dementia or MCI stages of AD may be too late in the disease process such that progressive neuronal loss and irreversible cognitive impairment may have already occurred (Jessen et al., 2014).Thus, conceptualization and investigation of the pre-MCI (i.e., preclinical) stage of AD are needed to define target populations for interventions at a stage of only mild neuronal damage and with still sufficient functional compensation (Sperling et al., 2011).

Because age is the biggest risk factor of late onset AD, changes in cognition, as functions of age, may be a combined effect of normal aging, AD-related pathology, and/or other neurodegenerative diseases. These cognitive changes are more pronounced in persons who are on a trajectory toward dementia, but this is difficult to distinguish from normal aging early in the process and methods are needed to improve early detection of cognitive decline (Toepper, 2017). Early detection and measurement of accelerating cognitive decline is important. A purpose of our research is to provide improved methods for recognizing when this optimal time window occurs by approximating it with a single point of transition in decrement, labelled a change point (CP).

Segmented linear regression or CP analysis constitutes a valuable framework for the investigation of the timing of cognitive decline and the rates of changes (Karr et al., 2018).

A systematic review of CP studies in dementia and AD shows consistent evidence for non-linear preclinical decline that varies by cognitive domain and demonstrates the utility of measuring cognitive functioning longitudinally in clinical practice, where observed declines in certain domains could serve as early predictors of a future Dementia diagnosis (Karr et al., 2018). An understanding of the empirical research on longitudinal cognitive change can be informative when differentiating AD from normal cognitive aging and potentially different dementia types.

Most previous CP methods align participants by anchoring them at time of diagnosis and then examining trajectories of variables of interest retrospectively for time points of change *prior to* clinical diagnosis. CPs have been detected at 15.5 years (Williams et al., 2020), 9.6 years (Thorvaldsson et al., 2011), 7.3 years (Grober et al., 2019), 6.8 year (Laukka et al., 2012), 6.6 years (Grober et al., 2019), 5.1 years (Hall et al., 2000) and 2.9 years (Grober et al., 2008) in different cognitive domains prior to AD diagnosis respectively. However, time to diagnosis is obviously not available for clinical use. Hall et al. (Hall et al., 2003) used both chronological age and time to dementia diagnosis to compare age-associated memory decline to disease-associated memory decline, but they only included people who developed dementia. Some studies used cross-sectional data (Luo et al., 2020) or retrospective data (Younes et al., 2019) to determine the ordering of changes in AD biomarkers among cognitively normal individuals, but these studies didn't allow biomarker CP to vary between participants. A method that can be applied prior to the onset of dementia to characterize accelerated preclinical decline with random CPs would thus be highly valuable when time to dementia is unknown.

Standard segmented linear regression is limited by the fact that the CP is modelled at the same time point for all individuals. This can be reasonable when using time to diagnosis as an explanatory variable but is inappropriate when modelling cognitive decline versus age. The CP in AD is highly variable across individuals and can occur anywhere between middle and extreme old age or it may not manifest itself at all. Several authors have, therefore, extended the model to treat the CP as a random effect parameter, thereby, allowing individuals to have distinct CPs (Brilleman et al., 2017; Dominicus et al., 2008; Hall et al., 2003; Jacqmin-Gadda et al., 2006; Segalas et al., 2019; Terrera, Van Den Hout, et al., 2011; van den Hout et al., 2011; Yang & Gao, 2013; Yu et al., 2012). The use of a random CP has the advantage of increasing model flexibility and is, therefore, likely to improve model fit without major alteration of parameter interpretation. Such models provide useful insights when the person-specific timing of the CP is of intrinsic interest, for example, in estimating the onset of cognitive decline in the elderly (Dominicus et al., 2008; Terrera, van den Hout, et al., 2011). However, the conditional distribution of the response variable in these models is assumed to be normal and does not allow us to flexibly model the entire distribution of the cognitive bent line curves at each age. For both clinical and research purposes, quantiles yield more information (e.g., those with high cognitive reserve can be defined as being above specified quantiles, and those in a low quantile can be considered at high risk and eligible for treatment and/or enrolment in a clinical trial). To avoid this shortcoming, Li et al. (C. Li et al., 2015) proposed a quantile regression CP model to describe changes in cognitive function in the preclinical and prodromal stages of AD. However, they assumed a fixed CP and modelled time until diagnosis.

We develop a subject-level bent-line regression model relating cognition to age, which then allows us to and indirectly recover population distributions and their associated quantiles. The latter, being mixtures of individual curves, will typically not be bent lines (see Fig. 2).

Several features in prospective studies of cognition versus age require flexibility when developing subject-specific CP models. First, we have repeated cognition measures and need to account for within-subject dependence at different times. Second, most subjects might not progress to AD or other dementia and their cognition scores versus age might tend to have very little decline and no observed CP. Third, changes in cognition leading to AD are directional and we expect greater decline in cognition after the CP. A model which accounts for random effects, a lack of CPs among some subjects, and constraints on the possible values of some parameters is difficult to fit in a typical likelihood framework using a frequentist estimation technique, so using a Bayesian approach is preferable (Hall et al., 2003). Bayesian methods allow prior distributions which may reflect previous knowledge and with modern software, such as the computational tool Stan (Carpenter et al., 2017), allow great model flexibility. For example, we can summarize posterior quantile curves of cognition versus age without imposing restrictive distribution assumptions.

We are aware of two previous studies of AD which develop Bayesian bent line CP models for cognition versus age and not time to diagnosis. Yu et al. (2012) studied elderly subjects already diagnosed with AD or MCI and measured the time of the CP after enrollment in the study. Li, Benitez, and Neelon (2021) studied a subset of the same elderly population where about one third were cognitively normal and used a latent variable approach to categorize diagnosis. They used a similar parameterization as we do with a directional constraint on the change in slope after the CP, but with different computational tool and only apply their analysis to a randomly chosen subset of fewer than 200 subjects. We are interested in applying the model to a younger and larger cohort where all subjects are cognitively unimpaired upon enrollment and most remain so throughout the study. A suitable data set arises from the Wisconsin Registry for

Alzheimer's Prevention (WRAP). Here we examine a three-test global preclinical Alzheimer's composite (PACC3) (Donohue et al., 2014; Jonaitis et al., 2019). More details about the WRAP cohort are given in (Johnson et al., 2018; Sager et al., 2005); and in Section 5 below.

The following sections will describe the construction of our BAyesian Bent-Line Regression (BaBLR) model with a random CP and constraints on the longitudinal cognitive composite score as a function of age among non-demented individuals at baseline using a prospective study, WRAP, and computational methods in Stan. In Section 2, we will present how the proposed model can accommodate longitudinal cognitive measurements. Section 3 presents the computational methods. Prior distribution specification and posterior inference are also described. In Section 4, the proposed model will be evaluated using simulation studies and Section 5 applies the proposed method to the WRAP dataset. Simulation studies mimic our data application both in terms of the number of subjects and the number of times each longitudinal measurement is observed. We will then close with a discussion of results and future work in Section 6.

## 2 BAyesian Bent-Line Regression model for longitudinal data

In this section, we propose a Bayesian constrained bent line regression with a random CP model for longitudinal data that involves four random effects (CPs, pre-change slopes, the differences between pre- and post-change slopes, and the expected response at CPs), which are modelled to vary among individuals around central population values. We do not constrain the slope before the CP since its pattern varies. However, the difference between pre- and post-change slopes is constrained to be negative, as we expect a more precipitous decline following the CP (H. Li et al., 2021). This approach may be used to analyse subtle longitudinal change; it is a flexible technique that is appropriate to use to study heterogeneity in change patterns. However, more complicated patterns of decline than a bent-line model may be more realistic. In practice, each subject is only measured a few times and this simple model is adequate to capture the primary feature of cognitive decline we seek. We should be cautious, however, in assuming linear trends to extend far beyond the extent of the measured data.

Let $y_{ij} = y_{ij}(t_{ij})$ denote the *j*th observed continuous outcome measurement taken for the *i*th participant ($i = 1,\ldots, N$) at some time points $t_{ij}$ ($j = 1, \ldots, n_i$) where $n_i$ is the number of measurements for subject $i$ and $N$ is the number of subjects.

We model the observed outcome measurements using a random CP mixed-effects model of the form

$$y_{ij} = \beta_{1i} + \beta_{2i}(t_{ij} - \omega_i)I(t_{ij} \leq \omega_i) + (\beta_{2i} + \beta_{3i})(t_{ij} - \omega_i)I(t_{ij} > \omega_i) + \varepsilon_{ij} \quad (1)$$

where $I$ is the indicator function, $\omega_i$ is the individual-specific CP, $\beta_{1i}$ is the individual-specific intercept denoting the expected value of the cognitive test score at the CP, $\beta_{2i}$ is the individual-specific linear slope before the CP (pre-change slope), and $\beta_{3i}$ is the individual-specific *difference* between pre- and post-change slope difference). We elect to model the slope difference rather than the post-slope directly in order to facilitate a sign constraint on this difference. We assume that the slope decrement $\beta_{3i}$ is less than or equal to zero, so that cognition declines more rapidly after the CP than before. Note that the model allows for cognition to continue to increase after the CP, but only with a smaller slope than before, although we expect most individuals will exhibit an actual negative slope. The individual-specific random parameters $\beta_{1i}$, $\beta_{2i}$, $\beta_{3i}$, $\omega_i$ can be further specified as

$$\begin{bmatrix} \beta_{1i} \\ \beta_{2i} \\ \beta_{3i} \\ \omega_i \end{bmatrix} = \begin{bmatrix} \beta_{10} \\ \beta_{20} \\ \beta_{30} \\ \omega_0 \end{bmatrix} + \begin{bmatrix} u_{1i} \\ u_{2i} \\ u_{3i} \\ u_{4i} \end{bmatrix} \quad (2)$$

such that $\beta_{10}$, $\beta_{20}$ and $\beta_{30}$ represent the fixed (population average) intercept, pre-change slope, and the slope *difference* between post-change and pre-change slope parameters, $\omega_0$ represents the fixed (population average) CP age, and $u_{1i}, u_{2i}, u_{3i}$ and $u_{4i}$ are the individual-level random effects (or deviations from the population average) associated with those parameters. The calculation of the maximum likelihood estimator subject to constraints is often difficult (Marchand & Strawderman, 2004). Hence, we use Bayesian methods for parameter estimation, since order constraints can be readily incorporated into the prior distributions for the model parameters. The term $\varepsilon_{ij}$ accounts for individual measurement noise.

We illustrate the model in Fig. 1.

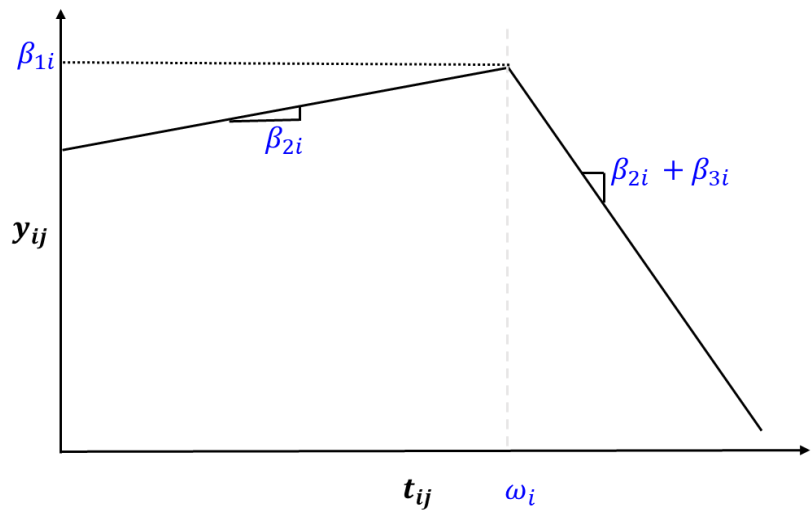

Fig. 1: BAyesian Bent-Line Regression (BaBLR) model illustration plot.

The parameters in this figure have the following interpretation: $\beta_{1i}, \beta_{2i}, \beta_{3i}, \omega_i$ denote the expected values for the *i*th subject.

## 3 Bayesian analysis

### 3.1 Summary of HMC algorithm

The BaBLR is fit using Hamiltonian Monte Carlo (HMC) via the freely available software Stan (Team, 2014). This method adaptively tunes the sampling process during a warm-up phase and can efficiently sample from posterior distributions with correlated parameters which is often the case for the cognitive model (Annis et al., 2017) and both built-in and user defined distributions can be used. The convergence of the HMC implementation was assessed using several available tools within the Stan software (Stan Development Team, 2020). We used R Version 3.6.2 for pre-processing of data and for post-processing and analysis of the MCMC samples (R Core Team, 2021). We interface with Stan from R using the RStan package (Team, 2014).

We ran 4 independent chains, each with 5000 iterations for warmup and 5000 iterations for inference. We modified the control values (e.g. increasing the target average proposal acceptance probability during Stan's adaptation period) to obtain more efficient sampling. we examined trace plots for each parameter, confirmed that autocorrelations were quite small after short lag times, and observed that the potential scale reduction statistic values $\hat{R}$ (Gelman & Rubin, 1992) were less than 1.05 for all parameters. We note that slow mixing of the pre-change slope required substantially longer chains using the WRAP data with the PACC3 outcome variable than we needed with simulated data other outcome variables from WRAP or other data sets for unknown reasons. The supplement contains the code for fitting BaBLR model.

*3.2 Prior distribution selection*

In the Bayesian framework, we need to specify the values of the hyperparameters of the prior distributions to compute the posterior distributions of the parameters. Following Gelman et al. (2006), we assume weakly informative prior distributions for all parameters. In our general model, we use normal prior distributions with random standard deviations for the population-level slopes and change point and for the individual-level random effects. We use a half-normal prior distribution for the difference in slopes. All scale parameters have half-Cauchy prior distributions. The likelihood uses a zero-mean normal model for the difference between individual outcomes and their expected value, also using a half-Cauchy prior distribution on the scale parameter. Thus: (i) the prior distributions for fixed and random effects parameters are taken to be independent normal or half-normal distributions, with details as follows

$$\beta_{k0} \sim N\left(\mu_{\beta k0}, \sigma_{\beta k0}\right) \text{ for } k = 1, 2, \omega$$

$$\beta_{30} \sim \text{Half-Normal}\left(\mu_{\beta 30}, \sigma_{\beta 30}\right)$$

$$u_{ki} \sim N\left(\mu_{uk}, \sigma_{uk}\right) \text{ for } k = 1,2,3,4 ;$$

(ii) For the scale parameters, we use a half-Cauchy distribution with location $\mu_{\sigma uk}$ and scale $\sigma_{\sigma uk}$ for $k = 1,2,3,4$; (iii) We use a half-Cauchy distribution with location $\mu_\varepsilon$ and scale $\sigma_\varepsilon$ for the residual standard deviation of the observed measurements $\sigma_y$. Note that $\sigma$ represents a scale parameter in the unit of measurement in each case. When the distribution is Normal, it is the standard deviation.

## 4 Simulation study

In this section we conduct two simulation studies to evaluate the finite sample performance of the proposed model. To test the implementation of the model, we conduct simulation 1 to verify that when data are simulated from the proposed model posterior estimates are appropriately recovered from the Bayesian analysis. For simplicity, we modelled the joint distribution of random effects for each subject and the unknown population level parameters as collections of independent random variables a priori, allowing the posterior sampling process to detect any associations among these variables. However, in practice, the individual-level random effects are not independent. For example, an individual with a higher than average mean response value at the change point might also tend to have a lower slope before the change point. Hence, to see if we can recover reasonable estimates of correlation between these random effects in the posterior distribution when present in the true model, even if the prior distribution models mutual independence, we conduct simulation 2. Simulated correlation values are based on observed WRAP study data. We use default control values for running Stan except increased adapt_delta and max_treedepth. In the supplement, we show the convergence results from our simulation study. In brief, we found that the four chains mixed very well, with $\hat{R} < 1.05$ for all parameters and low autocorrelations were.

*4.1 Simulation 1*

We generated datasets based on the proposed model (1) with $i = 1, 2, \ldots, n$, $j = 1, 2, \cdots, n_i$. We made assumptions similar to the preliminary results using our proposed model (1) from the WRAP data. For each individual $i$, we generated the number of biennial visits, $n_i$, from a discrete uniform U[3, 7] distribution. The time scale was age, ranging between 40 and 85 yrs. The details of prior distributions are as follows (Table 1):

Table 1. Simulation details for estimating model parameters

| Parameter | Value | k |
|---|---|---|
| $\mu_{\beta k0}$ | 0 | 1,2,3 |
|  | 10 | $\omega$ |
| $\mu_{uk}$ | 0 | 1,2,3,4 |
| $\mu_{\sigma\beta k0}$ | 0 | 1,2,3, $\omega$ |
| $\mu_{\sigma uk}$ | 0 | 1,2,3,4 |
| $\sigma_{\beta k0}$ | 10 | 1,4 |
|  | 1 | 2 |
|  | 5 | 3 |
| $\sigma_{\sigma uk}$ | 10 | 1,4 |
|  | 1 | 2 |
|  | 5 | 3 |
| $\mu_\varepsilon$ | 0 |  |
| $\sigma_\varepsilon$ | 10 |  |

The mean outcome at CP, slope before CP, and slope difference before and after CP and CP and the standard deviation of these parameters with the true values were shown in table 2. Three sample sizes $n$=100, $n$=500, and $n$=1000 were considered. We used mean bias (BIAS, across the 100 simulated datasets) and 95% coverage between Bayesian posterior estimates and their true population values from the preliminary results using our proposed model (1) from the WRAP data to assess the empirical performance of the parameter estimates. The simulation results were presented in Table 2. In all cases, we used 4 chains, 5000 warm-ups, and follow-up 5000 iterations.

As shown in Table 2, the average of the standard error estimates is slightly smaller than the empirical standard error for the CP estimator in every case. As a result, the coverage rate of every credible interval for the CP is lower than the nominal level of 95%. Our simulation showed that as the sample size increases the estimate of theoretical standard error approaches the empirical standard error and the empirical coverage probability of credible interval goes up towards 95%.

We also could estimate the distribution of the outcome trajectory. In Figure 2, we selected at random a single simulated data set of 100 subjects and plot the posterior quantile estimate of individual outcome trajectories versus age. We could see that while each individual trajectory follows the bent-line model, due to variation in the time of CPs among individuals, the population-level quantile curves are typically not bent lines and do not have sharply defined change points. We can, however, describe the general location in time where there is the strongest rate of change. Furthermore, we see evidence that the strongest rate of change occurs between 65 and 70 years at the 0.1 quantile, and conversely after 80 years at the 0.9.

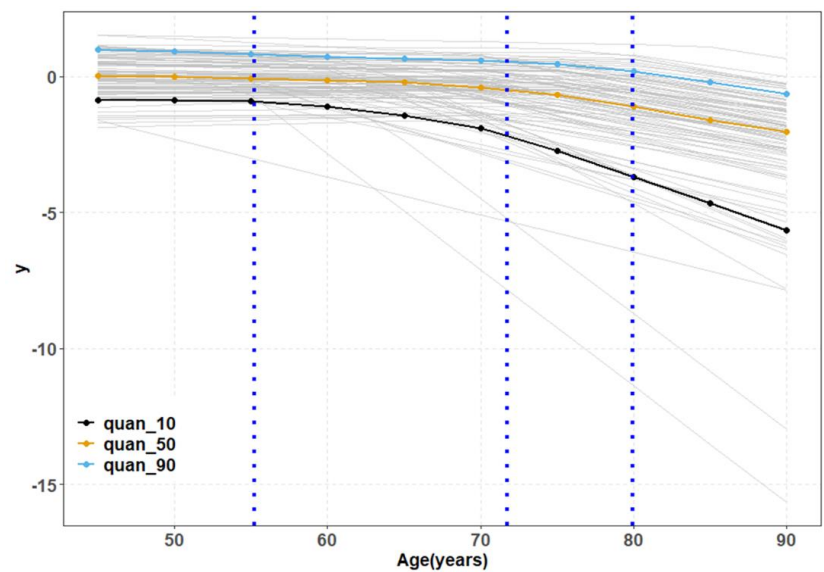

Fig. 2. The posterior quantile estimate of individual outcome trajectories versus age. The grey solid lines indicate the posterior median estimate of individual outcome trajectories, the colored solid lines are the 0.1, 0.5, and 0.9 quantiles, the blue vertical dotted line is the posterior median estimation of population average CP (fixed effects) with 95% credible interval (a single simulated data set of 100 subjects).

*4.2 Simulation 2*

Although we assume the four individual-specific random parameters $\beta_{1i}, \beta_{2i}, \beta_{3i}, \omega_i$ are independent, it is biologically plausible for there to be correlation. For example, when an individual has a higher outcome value at their CP, it is plausible that the rate of decline prior to the CP is less. This section assessed whether the model could recover the correlation between the random effects. We simulated the data with a true correlation matrix similar to preliminary results using our proposed model (1) from the WRAP data and used the uncorrelated prior distribution in the BaBLR model. The number of simulation runs was 30. The simulation results were presented in Table 3.

Table 3. Simulation results for estimating the correlation matrix

| Parameter | True values | Estimated mean (SE) | Bias |
|---|---|---|---|
| $\rho_{\sigma_{u_1}\sigma_{u_2}}$ | 0.807 | 0.615 (0.009) | -0.192 |
| $\rho_{\sigma_{u_1}\sigma_{u_3}}$ | 0.160 | 0.197 (0.024) | 0.037 |
| $\rho_{\sigma_{u_1}\sigma_{u_4}}$ | -0.553 | -0.597 (0.016) | -0.044 |
| $\rho_{\sigma_{u_2}\sigma_{u_3}}$ | 0.077 | 0.099 (0.020) | 0.022 |
| $\rho_{\sigma_{u_2}\sigma_{u_4}}$ | -0.423 | -0.518 (0.013) | -0.095 |
| $\rho_{\sigma_{u_3}\sigma_{u_4}}$ | -0.404 | -0.496 (0.021) | -0.092 |

The density plot of the posterior estimate correlation matrix is shown in Figure 3. The true correlation was almost in the middle of the posterior distribution for most of the parameter pairs, but the estimation is not good for the correlation between estimated outcome at CP and slope before CP, slope before CP and CP. As the association between the slope before the CP and the outcome value at the CP is not related to a question of substantial scientific interest in our case study, we are not concerned with the posterior distribution underestimating this relatively strong correlation. Moderate and weak correlations are recovered well. However, for different applications where accurate posterior estimation of potentially strong correlations is important, extending the model to include a more informative prior distribution for these correlations may be merited.

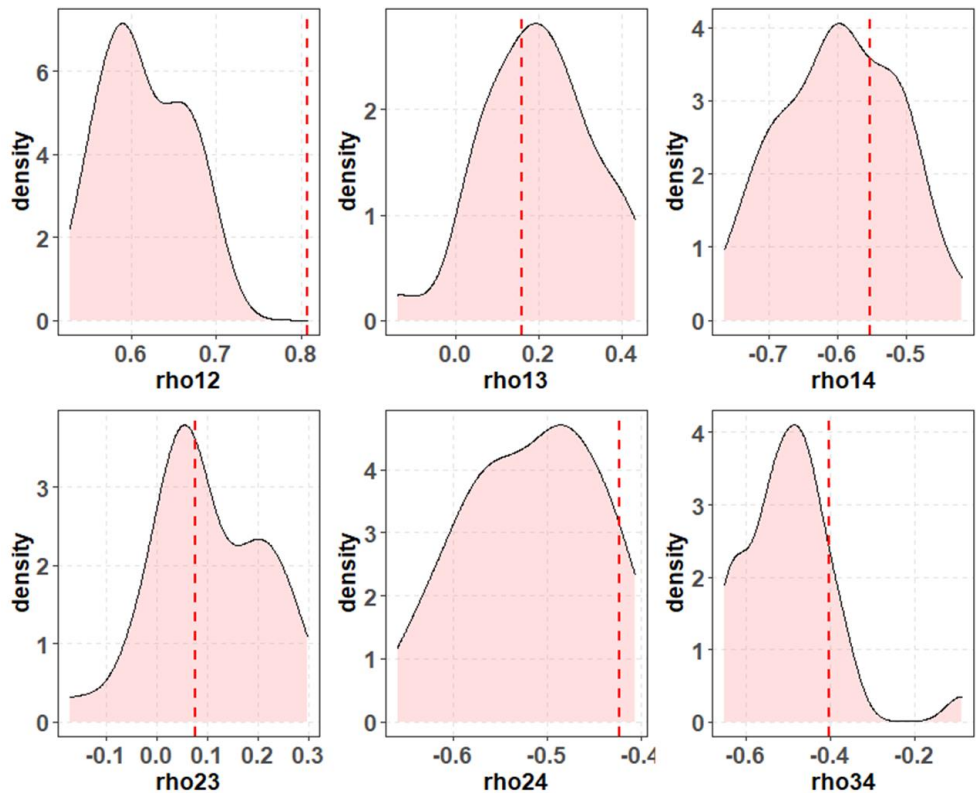

Fig. 3. The density plot of posterior estimate correlation matrix between estimate outcomes at CP (1), slope before CP (2), slope difference before and after CP (3), and CP (4). The red vertical dotted line is the true correlation.

Table 2. Simulation results for estimating model parameters

| Parameter | True values | N = 100 | | | N = 500 | | | N = 1000 | | |
|---|---|---|---|---|---|---|---|---|---|---|
| | | Estimated mean (SE) | Bias | 95% Cov* | Estimated mean (SE) | Bias | 95% Cov | Estimated mean (SE) | Bias | 95% Cov |
| $\beta_{10}$ | -0.0059 | -0.035 (0.009) | -0.029 | 0.94 | -0.014 (0.011) | -0.0083 | 0.90 | -0.023 (0.009) | -0.017 | 0.92 |
| $\beta_{20}$ | -0.0052 | -0.0056 (0.0004) | -0.0004 | 0.97 | -0.0052 (0.0003) | -0.0001 | 0.95 | -0.0056 (0.0003) | -0.00044 | 0.99 |
| $\beta_{30}$ | -0.0085 | -0.030 (0.001) | -0.021 | 0.98 | -0.018 (0.002) | -0.009 | 0.99 | -0.013 (0.002) | -0.005 | 0.99 |
| $\omega_0$ | 10 | 11.4 (0.22) | 1.4 | 0.93 | 11.4 (0.23) | 1.4 | 0.90 | 10.4 (0.19) | 0.4 | 0.99 |
| $\sigma_y$ | 0.30 | 0.30 (0.002) | -0.00007 | 0.95 | 0.30 (0.0015) | -0.002 | 0.90 | 0.30 (0.0012) | -0.003 | 0.92 |
| $\sigma_{u_1}$ | 0.64 | 0.62 (0.006) | -0.022 | 0.96 | 0.63 (0.005) | -0.009 | 0.99 | 0.63 (0.003) | -0.008 | 0.99 |
| $\sigma_{u_2}$ | 0.02 | 0.02 (0.0002) | -0.0081 | 0.96 | 0.014 (0.0002) | 0.002 | 0.96 | 0.013 (0.0003) | -0.0073 | 0.96 |
| $\sigma_{u_3}$ | 0.15 | 0.14 (0.002) | -0.0067 | 0.94 | 0.15 (0.002) | -0.005 | 0.95 | 0.15 (0.002) | -0.005 | 0.95 |
| $\sigma_{u_4}$ | 10 | 10.5 (0.14) | 0.50 | 0.98 | 10.6 (0.16) | 0.60 | 0.90 | 10.2 (0.14) | 0.20 | 0.99 |

* in 100 trials with a 0.95 success probability, there is over a 95% chance that the sample coverage percentage falls between the 0.025 and 0.975 quantiles of 90% and 99%, respectively.

# 5  Application to the assessment of cognitive function in a longitudinal cohort study

## 5.1  Study design

The WRAP is a longitudinal observational cohort study enriched with persons having a parental history (PH) of probable Alzheimer's disease (AD) dementia. The WRAP cohort includes neuropsychological data from 1606 participants who enrolled at midlife and were free of dementia at baseline (Johnson et al., 2018; Sager et al., 2005). Enrolment began in 2001; follow-up assessments with second-wave assessments were conducted approximately 2-4 years after baseline and all subsequent wave follow-up visits were conducted at approximately 2-year intervals, with median follow up of 9 years. The biannual, comprehensive neuropsychological battery covered all major cognitive domains (see Johnson et al., 2018 for a complete description). Cognitive status was determined at each visit was classified as CU-S (cognitively unimpaired – standard), CU-D (cognitively unimpaired – declining), MCI (mild cognitive impairment), non-MCI impaired (i.e., impairment such as that associated with the presence of a learning disability), and Dementia (Langhough Koscik et al. 2021).

## 5.2  Selection of Participants

Participants that had at least three PACC3 scores are included in this study (n(%) = 1068 (66.5%) of 1606 after exclusions).

A summary of the characteristics of the sample and the overall data are presented in Table 4. Of 1068 mean (SD) age at the first cognitive assessment was 58.4(6.4) [range 40.2–73.6] with mean (SD) = 8.2 (2.2) years between the first and last cognitive assessment. At the last cognitive visit, consensus conference review of data indicated that 908 (85.0%) were cognitively unimpaired-stable, 123 (11.5%) were CU-D, 33 (3.1%) were MCI, and 4 (0.4%) had progressed to dementia.

Table 4. Characteristics of the Overall and sample for analysis

|  | Overall | Analyses data |
|---|---|---|
| N | 1606 | 1068 |
| Age at enrolment (years; mean (SD)) | 54.49 (6.8) | 54.14 (6.5) |
| Age at first PACC3 (years; mean (SD))[a] | 58.74 (6.6) | 58.40 (6.4) |
| Female (%) | 1135 (70.7) | 738 ( 69.1) |
| *APOE* e4 carriers (%) | 609 (39.6) | 407 ( 38.1) |
| Family History (%) | 1178 (73.3) | 795 ( 74.4) |
| College degree (%) | 932 (58.0) | 677 ( 63.4) |
| WRAT3 Reading (mean (SD)) | 104.60 (10.3) | 107.81 (8.7) |
| Cognitive Status (at enrollment) (%) | | |
| CU-S | 1415 (88.1) | 965 ( 90.3) |
| CU-D | 189 (11.8) | 102 ( 9.6) |
| MCI | 2 ( 0.1) | 1 ( 0.1) |

*Overall: Participants who were enrolled with at least a baseline visit at the time of these analyses; Analyses data: Participants who met enrolment criteria.

Abbreviations: College degree, Education years>=16; WRAT3, wide range achievement test (third edition); PACC3, global cognitive composite 3; CU-S, cognitively unimpaired – standard; CU-D, cognitively unimpaired – declining; MCI, mild cognitive impairment.

[a] The sample size for the first column is N = 1390, second column is N = 1068.

### *5.3 Model Building*

Episodic memory quantiles were modelled assuming a piecewise linear trajectory with a CP at the time which the decline of episodic memory quantile accelerated (C. Li et al., 2015). The inspection of spaghetti plots of PACC3 scores against age (Figure 4) suggested that the CP might be different for different individuals on the scale of age, and the CP's and slopes are different across cognitive statuses (as shown by color coding a subset of the lines by last cognitive status).

We fit Equation 1 to the 1068 subjects in WRAP using cognitive composite PACC3 score as the outcome and age in years at the time scale. We assume the PACC3 decline rate after CP is faster than it is before CP ($\beta_{3i} < 0$) because we are not interested in when the slope after CP increases, as such a change is inconsistent with the onset of dementia and may be better explained as learning from repeated assessment (i.e., practice effects). Also, we focus on sporadic AD, so we assume the CP could not occur before the age of 40. We used the same prior distributions as in the simulation 1 (Table 1) for all the parameters with the exception of $\sigma_{u2} \sim \text{lognormal}(0, 0.2)$ due to poor mixing behavior when using a half-Cauchy prior distribution for this parameter. We discuss in more detail when we describe the sensitivity analysis. The computational burden to run 10,000 HMC iterations took approximately 10h on a machine with an AMD Ryzen 7 1800X CPU @3.85GHz processor, 8 cores, and 64GB RAM.

In the supplement, we showed the convergence results from the application. In brief, we found the four chains mixed very well, $\hat{R} < 1.05$ for all parameters and autocorrelations were low.

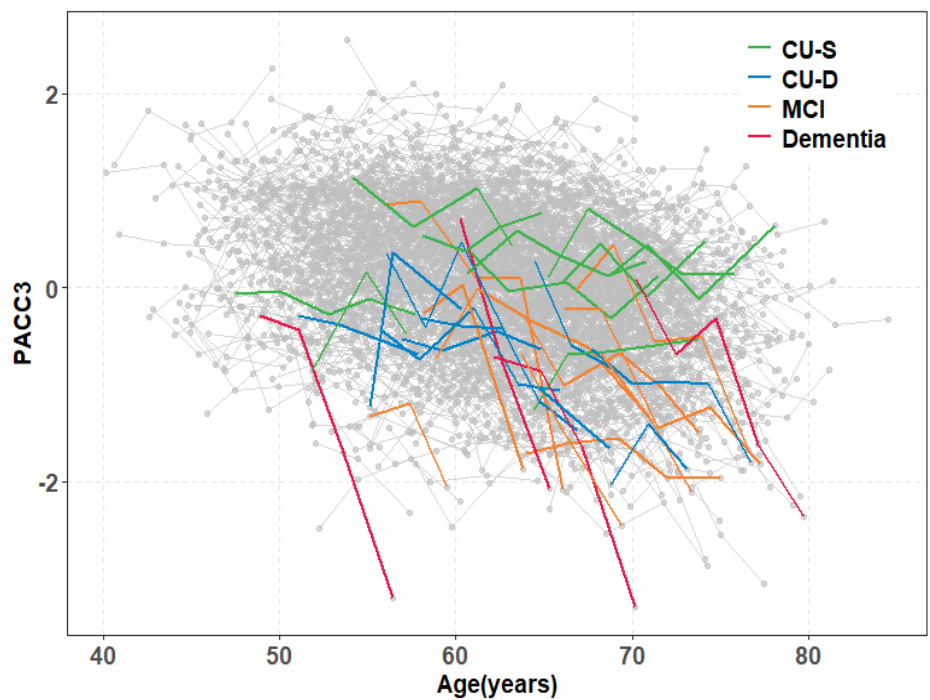

Fig. 4. Spaghetti plot of individual PACC3 trajectories versus age by last cognitive status (Total **n = 34**; randomly chose 10 CU-S, 10 CU-D, 10 MCI and all with Dementia). The grey dots and solid lines indicate observed data (n = 1068). Abbreviations: PACC3, global cognitive composite 3; CU-S, cognitively unimpaired – standard; CU-D, cognitively unimpaired – declining; MCI, mild cognitive impairment.

*5.4 Analysis Results*

Using data from the full PACC3 sample, we used our mixed model in which the CP is allowed to vary across participants. This feature is crucial for real-world application. BaBLR successfully estimated fixed and random parameters. Details of the posterior median estimates and 95% credible intervals are in column 1 of Table 5. The fit of the BaBLR was illustrated in Figs. 5 and 6. Figure 5 shows the posterior quantile estimate of individual outcome trajectories versus age. The lower quantile has earlier CPs and a faster decline after CPs; the upper quantile has later CPs and a slower decline after CPs. We display the observed and estimated trajectories for 2 selected subjects from each of these CP clusters in Figure 6: the estimated median CP is at the beginning or before the first visit; during visits, and at their last visit. The fit is good with individual predicted trajectories matching the observed marker values. The Fig 6a) subject 1's CP was estimated to be 3 years prior to first cognitive assessment, with MCI observed at the last visit (10 years after CP). Fig 6b) show the data and estimates for a subject without any clinical diagnoses. People who have MCI or dementia are likely to have a CP at a younger age, and most people with normal cognition may have CP at a very late age, and the 0.5, 0.8, and 0.95 quantile intervals are wider. As shown in Figure 5, the range of extrapolation at older ages is wider, therefore, the uncertainty is bigger.

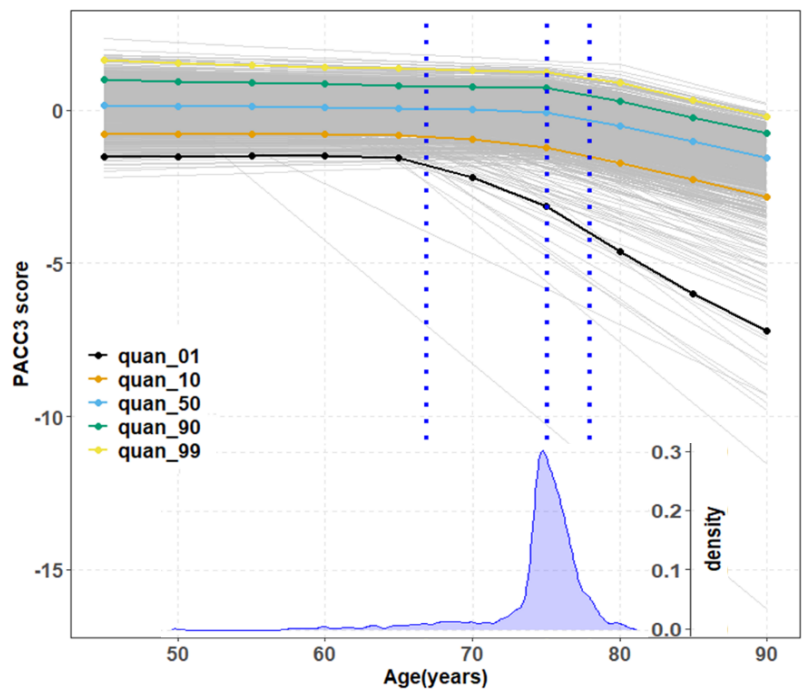

Fig. 5. The posterior quantile estimate of individual outcome trajectories versus age. The grey solid lines indicate the posterior median estimate of individual outcome trajectories, the colored solid lines are quantiles 0.01 to 0.99, the blue vertical dotted line is the posterior median estimate of CP with 95% credible interval, and the density plot is the posterior estimate of CP.

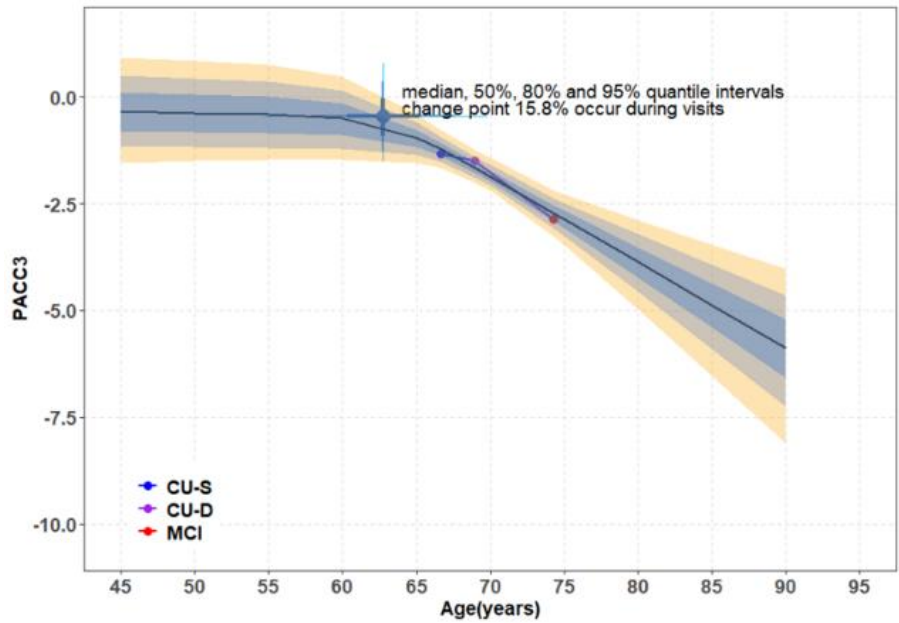

(a)

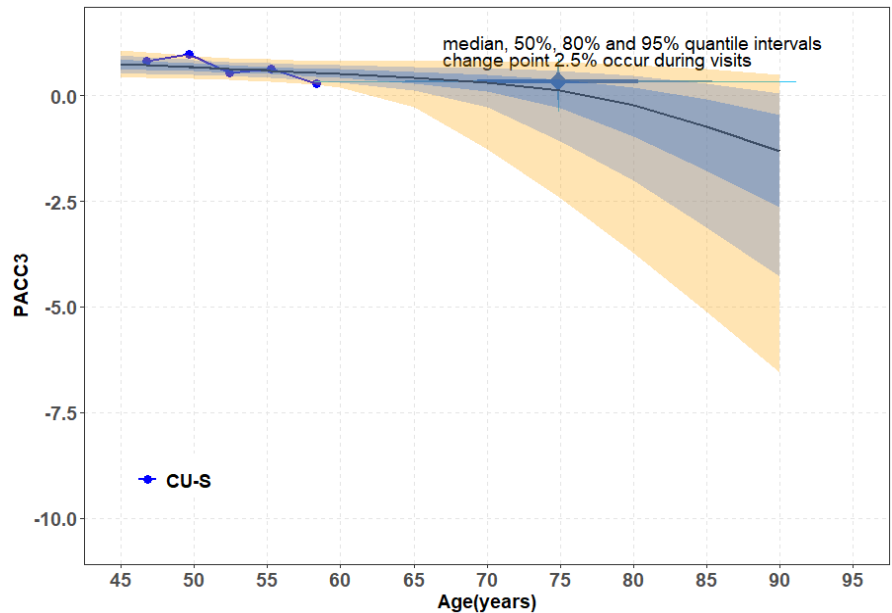

(b)

Fig. 6. The fit of the BaBLR for PACC3 for two participants. The colored dots and solid lines by cognitive status are individual observations (blue: CU-S; purple: CU-D; red: MCI), the black solid lines are predicted trajectories, the colored ribbons are 0.5, 0.8, and 0.95 quantile intervals of these predicted trajectories, and the light blue dot is the median CP and estimate PACC3 at CP with 0.5, 0.8 and 0.95 quantile intervals. 6a shows the estimated median CP is before the first visit and 6b shows the estimated median CP is after their last visit.

Caution must be taken regarding sensitivity to the choice of the prior distribution, identifiability, and goodness of fit. Two sensitivity analyses were performed to estimate whether the results are driven by priors. First, we chose different half normal, log normal, and student t distributions to illustrate why we choose this prior and why it is better.

Table 5. Estimates and 95% credible intervals for the Parameters of the Model with different priors for scale parameters

| Parameter | Model 1 (main model) | Model 2 (half_cauchy) | Model 3 (half_normal) | Model 4 (half_t student) |
|---|---|---|---|---|
| $\beta_{10}$ | 0.0015 [-0.049,0.075] | -0.00047 [-0.070,0.055] | -0.0023 [-0.082,0.053] | -0.0022 [-0.085,0.057] |
| $\beta_{20}$ | -0.0045 [-0.0076,-0.00051] | -0.0053 [-0.0084,-0.0022] | -0.0054 [-0.0087,-0.0023] | -0.0054 [-0.0087,-0.0022] |
| $\beta_{30}$ | -0.0028 [-0.024,-0.00001] | -0.0034 [-0.030,-0.00001] | -0.0043 [-0.037,-0.00001] | -0.0040 [-0.036,-0.00001] |
| $\omega_0$ | 74.95 [72.61,76.64] | 75.21 [73.80,78.24] | 75.91 [73.27,79.13] | 75.89 [73.25,79.05] |
| $\sigma y$ | 0.30 [0.29,0.30] | 0.30 [0.29,0.31] | 0.30 [0.29,0.31] | 0.30 [0.29,0.31] |
| $\sigma u_1$ | 0.80 [0.77,0.82] | 0.81 [0.79,0.83] | 0.81 [0.79,0.83] | 0.81 [0.79,0.83] |
| $\sigma u_2$ | 0.12 [0.11,0.13] | 0.036 [0.017,0.087] | 0.049 [0.014,0.096] | 0.053 [0.027,0.094] |
| $\sigma u_3$ | 0.38 [0.34,0.43] | 0.39 [0.35,0.44] | 0.39 [0.35,0.44] | 0.39 [0.35,0.45] |
| $\sigma u_4$ | 2.88 [2.63,3.15] | 2.93 [2.68,3.23] | 2.96 [2.68,3.27] | 2.95 [2.68,3.26] |

As shown in Table 5, all the parameter estimates are similar except $\sigma u_2$, the standard deviation of individual random slopes before CPs. There is a convergence problem for $\sigma u_2$ when we use half-Cauchy, half-normal, and half-t_student (Rhat > 1.05). We also compared the density plots of the fixed and random effects when we use the same priors for all parameters except $\sigma u_2 \sim$ lognormal (0, 0.2) in model 1 and $\sigma u_2 \sim$ half-Cauchy (0, 1) in model 2 (Fig. 7). The density curves are similar for models 1 and 2 but shifts to the left slightly for population intercept and slope before CPs and shifts to the right slightly for population CPs and the individual scale parameters in model 2. In practice, there is variation of PACC3 measurement at CP in population, but the difference between model 1 and 2 is 0.8±0.01, which is very small, and thus not clinically meaningful. The big difference is σu2, which curve is right-skewed and close to 0 in model 2, but the curve is an approximately normal distribution with a mean of 0.12 and a standard deviation of 0.007.

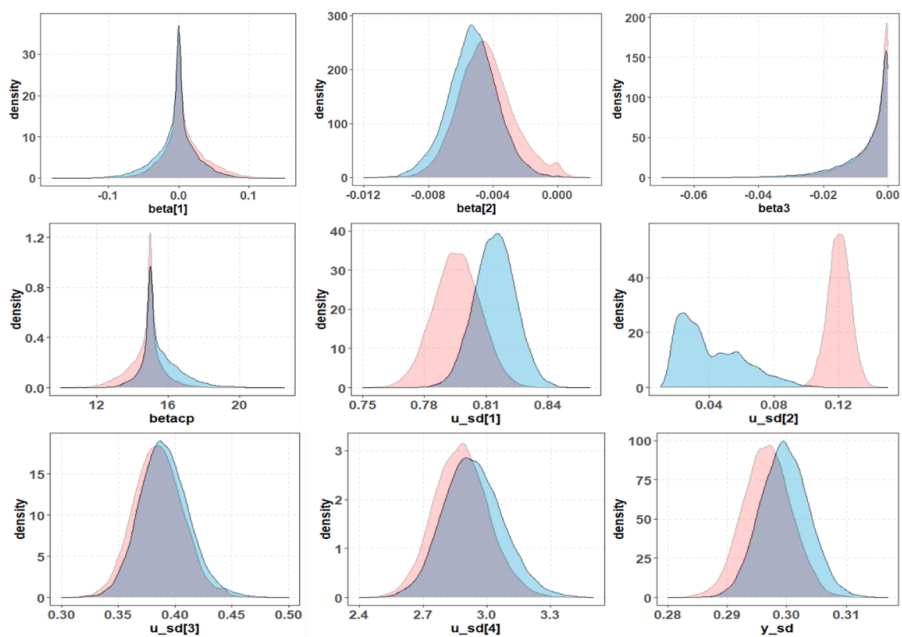

Fig. 7. The comparison of the BaBLR density plot for PACC3 using different priors. The pink color indicates model 1 using the prior u_sd[2] ~ lognormal (0, 0.2). The blue color indicates model 2 using the prior u_sd[2] ~ half-Cauchy (0, 1). Abbreviations: beta[1], the estimated population mean outcome at CP; beta[2], the estimated population mean slope before CP; beta3, the estimated population mean difference between before and after slope; betacp, the estimate population mean CP; u_sd[1], the estimate standard deviation individual outcome at CP; u_sd[2], the estimate standard deviation individual slope before CP; u_sd[3], the estimate standard deviation individual difference between before and after slope; u_sd[4], the estimate standard deviation individual CP; y_sd, the estimate error term.

Second, we chose different priors for the mean of the CP to look at whether the CPs are driven by the priors. The prior CP mean in the main results (Table 5, Model 1) is 70, here we chose 65, 75, and 80. The results were shown in Supplemental Table S1. As shown in Table S1, the results are similar, so the CPs are not driven by priors.

Validation was also performed for the BaBLR model. We randomly selected half individuals to leave out the last observations and look at how many last observations fall in .95 quantile of the predicted interval.

Let $Y = \mu + X, \mu \sim N(\mu_0, \sigma_0), X \sim N(0, \sigma_x)$,

$$P(Y \leq y) = E(P(Y \leq y | \mu)) \approx \frac{\sum_{i=1}^{n} \phi(\frac{y-\mu}{\sigma_x})}{n} \quad (4)$$

We extracted the posterior individual estimates from 20000 model runs to calculate the predicted outcome and used Eq. (4) to find the .025, .5, and .975 quantiles and then calculate the .95 quantile of estimated last observations for each individual. There is a linear relationship between the .5 quantile of the estimated last observation and the true last PACC3 for each individual (Fig. S1). Fig. S2 shows 99.3% of the last PACC3 falls in the .95 quantile of estimate at last observation, with 1 low true value, and 3 high true values.

There are several advantages of our model. First, compared with the fixed CP quantile model (C. Li et al., 2015), ours uses age as a time scale and studies subject-specific bent line CP models in which the within-subject dependence among data is taken into account through the incorporation of random effects, which is directly clinically useful. Second, compared with the random CP model in the frequentist method, our model does not have the difficulty of accounting for random CPs in a likelihood framework using a frequentist estimation technique. Third, compared with the Bayesian random CP mean model, which could be implemented using "brms" package (Bürkner, 2017), our model could estimate the distribution of individual trajectories. The distribution of cognitive trajectories by age is not normal, therefore, only mean and standard deviation is inefficient. In addition, our model adds a constrains to the slope difference between before and after CP, as cognitive decline progresses more rapidly following the change point (H. Li et al., 2021).

## 6 Discussion

We have proposed a BAyesian Bent-Line Regression model to estimate the timings of random CPs, and the trajectories for longitudinal cognitive data accounting for intra-subject correlation and applied it to a sample that was cognitively unimpaired at baseline cognition but at increased risk of progressing to MCI or dementia. Studies of MCI or dementia treatments usually attempt to identify those who are prodromal, before symptoms develop, but a global Clinical Dementia Rating (CDR) of 0.5 might not be sensitive to the earliest changes in cognition (Wada-Isoe et al., 2019). Instead, our research provided improved methods for quickly recognizing when this optimal time window occurs by approximating it with a CP. Also, our model allows subject-specific random effects to account for inter-individual variations. Although the trajectory parameters do not depend upon covariates in our example, they are easy incorporate. The performance of our model was studied through simulation. Results of the application to the WRAP study showed the timing and slopes of cognitive change vary greatly between individuals. Through the use of a random CP, our model provided the flexibility required to estimate the individual-specific timing of cognitive change for each participant, while also providing an estimate of the mean timing of cognitive change and the variability around that mean. The CP analyses presented here lead to several findings. First, the timing of CPs tends to precede clinical deterioration, i.e. people who have MCI or dementia might have CPs several years before diagnosis. Second, the CPs are variable, with their random effect having an estimated standard deviation of about 3 years. Third, people who are cognitively unimpaired at all visits might have estimated CPs at very late ages. Fourth, the cognitive trajectory (i.e., slope after CP) is flat in the highest quantile and changes earlier and faster in lower quantile.

There are at least three future research topics on this random CP mixed effect model. One is to model the effects of covariates (e.g. genetic markers) on the CP. This will lead to a unified model for the entire data rather than separate models for data sets separated by covariates that significantly affect the CP and/or its associated covariate's coefficients. Another direction is to investigate two random CPs in the model (e.g., to capture patterns that include plateaus in decline). It appears there are two CPs when people who developed MCI or dementia have earlier CPs. However, we might need a larger dataset to fit this model. The third research topic is the joint modelling of the random CPs and survival outcomes such as progression to dementia. The drawback of the current model is we don't consider any survival outcome, but we might need a longer follow-up to have enough survival outcomes.

Predicting individual cognitive trajectories and CPs provides an opportunity for early intervention by identifying subjects at high-risk of subsequent dementia at the right time for treatment and/or enrollment in a clinical trial.

**Acknowledgments**

We gratefully acknowledge our dedicated WRAP participants and the personnel from the study teams associated with all the grants contributing to this study's data.

**Supplementary Materials**

The supplemental materials contain additional details on sensitivity analysis, the convergence results for simulation and application study, R and Stan code simulating the dataset and implementing the methods. They augment further the data, methods, and results presented in the main article.

**Data Availability Statement**

All data and materials used within this study will be made available, upon reasonable request, to research groups wishing to reproduce/confirm our results.

**Disclosure Statement**

The authors report there are no competing interests to declare.

**Additional information**

**Funding**

This work is supported by NIH/NIA R01AG021155, RF1AG027161, P30AG062715, and Alzheimer's Association AARF-19-614533.

**References:**

Bäckman, Lars, Sari Jones, Anna Karin Berger, Erika Jonsson Laukka, and Brent J. Small. 2005. "Cognitive Impairment in Preclinical Alzheimer's Disease: A Meta-Analysis." *Neuropsychology* 19(4):520–31. doi: 10.1037/0894-4105.19.4.520.

Brilleman, Samuel L., Laura D. Howe, Rory Wolfe, and Kate Tilling. 2017. "Bayesian Piecewise Linear Mixed Models with a Random Change Point: An Application to BMI Rebound in Childhood." *Epidemiology* 28(6):827–33. doi: 10.1097/EDE.0000000000000723.

Budd Haeberlein, Samantha, P. S. Aisen, F. Barkhof, S. Chalkias, T. Chen, S. Cohen, G. Dent, O. Hansson, K. Harrison, C. von Hehn, T. Iwatsubo, C. Mallinckrodt, C. J. Mummery, K. K.

Muralidharan, I. Nestorov, L. Nisenbaum, R. Rajagovindan, L. Skordos, Y. Tian, C. H. van Dyck, B. Vellas, S. Wu, Y. Zhu, and A. Sandrock. 2022. "Two Randomized Phase 3 Studies of Aducanumab in Early Alzheimer's Disease." *Journal of Prevention of Alzheimer's Disease* 9(2):197–210. doi: 10.14283/JPAD.2022.30/TABLES/3.

Bürkner, Paul-Christian. 2017. "Brms: An R Package for Bayesian Multilevel Models Using Stan." *Journal of Statistical Software* 80(1 SE-Articles):1–28. doi: 10.18637/jss.v080.i01.

Dominicus, Annica, Samuli Ripatti, Nancy L. Pedersen, and Juni Palmgren. 2008. "A Random Change Point Model for Assessing Variability in Repeated Measures of Cognitive Function." *Statistics in Medicine* 27(27):5786–98. doi: 10.1002/sim.3380.

Dubois, Bruno, Howard H. Feldman, Claudia Jacova, Jeffrey L. Cummings, Steven T. DeKosky, Pascale Barberger-Gateau, André Delacourte, Giovanni Frisoni, Nick C. Fox, Douglas Galasko, Serge Gauthier, Harald Hampel, Gregory A. Jicha, Kenichi Meguro, John O'Brien, Florence Pasquier, Philippe Robert, Martin Rossor, Steven Salloway, Marie Sarazin, Leonardo C. de Souza, Yaakov Stern, Pieter J. Visser, and Philip Scheltens. 2010. "Revising the Definition of Alzheimer's Disease: A New Lexicon." *The Lancet. Neurology* 9(11):1118–27. doi: 10.1016/S1474-4422(10)70223-4.

Dubois, Bruno, Nicolas Villain, Giovanni B. Frisoni, Gil D. Rabinovici, Marwan Sabbagh, Stefano Cappa, Alexandre Bejanin, Stéphanie Bombois, Stéphane Epelbaum, Marc Teichmann, Marie Odile Habert, Agneta Nordberg, Kaj Blennow, Douglas Galasko, Yaakov Stern, Christopher C. Rowe, Stephen Salloway, Lon S. Schneider, Jeffrey L. Cummings, and Howard H. Feldman. 2021. "Clinical Diagnosis of Alzheimer's Disease: Recommendations of the International Working Group." *The Lancet Neurology* 20(6):484–96. doi: 10.1016/S1474-4422(21)00066-1.

Gelman, Andrew, and Donald B. Rubin. 1992. "Inference from Iterative Simulation Using Multiple Sequences." *Https://Doi.Org/10.1214/Ss/1177011136* 7(4):457–72. doi: 10.1214/SS/1177011136.

Grober, Ellen, Yang An, Richard B. Lipton, Claudia Kawas, and Susan M. Resnick. 2019. "Timing of Onset and Rate of Decline in Learning and Retention in the Pre-Dementia Phase of Alzheimer's Disease." *Journal of the International Neuropsychological Society : JINS* 25(7). doi: 10.1017/S1355617719000304.

Grober, Ellen, Charles B. Hall, Richard B. Lipton, Alan B. Zonderman, Susan M. Resnick, and Claudia Kawas. 2008. "Memory Impairment, Executive Dysfunction, and Intellectual Decline in Preclinical Alzheimer's Disease." *Journal of the International Neuropsychological Society : JINS* 14(2):266–78. doi: 10.1017/S1355617708080302.

Hall, Charles B., Richard B. Lipton, Martin Sliwinski, and Walter F. Stewart. 2000. "A Change Point Model for Estimating the Onset of Cognitive Decline in Preclinical Alzheimer's Disease." *Statistics in Medicine* 19(11-12):1555–66. doi: 10.1002/(SICI)1097-0258(20000615/30)19:11/12<1555::AID-SIM445>3.0.CO;2-3.

Hall, Charles B., Jun Ying, Lynn Kuo, and Richard B. Lipton. 2003. "Bayesian and Profile Likelihood Change Point Methods for Modeling Cognitive Function over Time." *Computational Statistics and Data Analysis* 42(1–2):91–109. doi: 10.1016/S0167-9473(02)00148-2.

van den Hout, Ardo, Graciela Muniz-Terrera, and Fiona E. Matthews. 2011. "Smooth Random Change Point Models." *Statistics in Medicine* 30(6):599–610. doi: 10.1002/sim.4127.

Jack, Clifford R., David A. Bennett, Kaj Blennow, Maria C. Carrillo, Billy Dunn, Samantha Budd Haeberlein, David M. Holtzman, William Jagust, Frank Jessen, Jason Karlawish, Enchi Liu, Jose Luis Molinuevo, Thomas Montine, Creighton Phelps, Katherine P. Rankin, Christopher C. Rowe, Philip Scheltens, Eric Siemers, Heather M. Snyder, Reisa Sperling, Cerise Elliott, Eliezer Masliah, Laurie Ryan, and Nina Silverberg. 2018. "NIA-AA Research Framework: Toward a Biological Definition of Alzheimer's Disease." *Alzheimer's & Dementia* 14(4):535–62. doi: 10.1016/J.JALZ.2018.02.018.

Jacqmin-Gadda, Hélène, Daniel Commenges, and Jean-François Dartigues. 2006. "Random Changepoint Model for Joint Modeling of Cognitive Decline." *Biometrics* 62(1):254–60. doi: 10.1111/j.1541-0420.2005.00443.x.

Jessen, Frank, Rebecca E. Amariglio, Martin Van Boxtel, Monique Breteler, Mathieu Ceccaldi, Gaël Chételat, Bruno Dubois, Carole Dufouil, Kathryn A. Ellis, Wiesje M. Van Der Flier, Lidia Glodzik, Argonde C. Van Harten, Mony J. De Leon, Pauline McHugh, Michelle M. Mielke, Jose Luis Molinuevo, Lisa Mosconi, Ricardo S. Osorio, Audrey Perrotin, Ronald C. Petersen, Laura A. Rabin, Lorena Rami, Barry Reisberg, Dorene M. Rentz, Perminder S. Sachdev, Vincent De La Sayette, Andrew J. Saykin, Philip Scheltens, Melanie B. Shulman, Melissa J. Slavin, Reisa A. Sperling, Robert Stewart, Olga Uspenskaya, Bruno Vellas, Pieter Jelle Visser, and Michael Wagner. 2014. "A Conceptual Framework for Research on Subjective Cognitive Decline in Preclinical Alzheimer's Disease." *Alzheimer's & Dementia : The Journal of the Alzheimer's Association* 10(6):844. doi: 10.1016/J.JALZ.2014.01.001.

Johnson, Sterling C., Rebecca L. Koscik, Erin M. Jonaitis, Lindsay R. Clark, Kimberly D. Mueller, Sara E. Berman, Barbara B. Bendlin, Corinne D. Engelman, Ozioma C. Okonkwo, Kirk J. Hogan, Sanjay Asthana, Cynthia M. Carlsson, Bruce P. Hermann, and Mark A. Sager. 2018. "The Wisconsin Registry for Alzheimer's Prevention: A Review of Findings and Current

Directions." *Alzheimer's & Dementia : Diagnosis, Assessment & Disease Monitoring* 10:130. doi: 10.1016/J.DADM.2017.11.007.

Jonaitis, Erin M., Rebecca L. Koscik, Lindsay R. Clark, Yue Ma, Tobey J. Betthauser, Sara E. Berman, Samantha L. Allison, Kimberly D. Mueller, Bruce P. Hermann, Carol A. Van Hulle, Bradley T. Christian, Barbara B. Bendlin, Kaj Blennow, Henrik Zetterberg, Cynthia M. Carlsson, Sanjay Asthana, and Sterling C. Johnson. 2019. "Measuring Longitudinal Cognition: Individual Tests versus Composites." *Alzheimer's & Dementia : Diagnosis, Assessment & Disease Monitoring* 11:74. doi: 10.1016/J.DADM.2018.11.006.

Karr, Justin E., Raquel B. Graham, Scott M. Hofer, and Graciela Muniz-Terrera. 2018. "When Does Cognitive Decline Begin? A Systematic Review of Change Point Studies on Accelerated Decline in Cognitive and Neurological Outcomes Preceding Mild Cognitive Impairment, Dementia, and Death HHS Public Access." *Psychol Aging* 33(2):195–218. doi: 10.1037/pag0000236.

Koscik, Rebecca L., Tobey J. Betthauser, Erin M. Jonaitis, Samantha L. Allison, Lindsay R. Clark, Bruce P. Hermann, Karly A. Cody, Jonathan W. Engle, Todd E. Barnhart, Charles K. Stone, Nathaniel A. Chin, Cynthia M. Carlsson, Sanjay Asthana, Bradley T. Christian, and Sterling C. Johnson. 2020. "Amyloid Duration Is Associated with Preclinical Cognitive Decline and Tau PET." *Alzheimer's & Dementia: Diagnosis, Assessment & Disease Monitoring* 12(1):e12007. doi: 10.1002/DAD2.12007.

Laukka, Erika J., Stuart W. S. MacDonald, Laura Fratiglioni, and Lars Bäckman. 2012. "Preclinical Cognitive Trajectories Differ for Alzheimer's Disease and Vascular Dementia." *Journal of the International Neuropsychological Society : JINS* 18(2):191–99. doi: 10.1017/S1355617711001718.

Li, Chenxi, N. Maritza Dowling, and Rick Chappell. 2015. "Quantile Regression with a Change-Point Model for Longitudinal Data: An Application to the Study of Cognitive Changes in Preclinical Alzheimer's Disease." *Biometrics* 71(3):625. doi: 10.1111/BIOM.12313.

Li, Hong, Andreana Benitez, and Brian Neelon. 2021. "A Bayesian Hierarchical Change Point Model with Parameter Constraints." *Statistical Methods in Medical Research* 30(1):316. doi: 10.1177/0962280220948097.

Livingston, Gill, Andrew Sommerlad, Vasiliki Orgeta, Sergi G. Costafreda, Jonathan Huntley, David Ames, Clive Ballard, Sube Banerjee, Alistair Burns, Jiska Cohen-Mansfield, Claudia Cooper, Nick Fox, Laura N. Gitlin, Robert Howard, Helen C. Kales, Eric B. Larson, Karen Ritchie, Kenneth Rockwood, Elizabeth L. Sampson, Quincy Samus, Lon S. Schneider, Geir Selbæk,

Linda Teri, and Naaheed Mukadam. 2017. "Dementia Prevention, Intervention, and Care." *Lancet (London, England)* 390(10113):2673–2734. doi: 10.1016/S0140-6736(17)31363-6.

Luo, Jingqin, Folasade Agboola, Elizabeth Grant, Colin L. Masters, Marilyn S. Albert, Sterling C. Johnson, Eric M. McDade, Jonathan Vöglein, Anne M. Fagan, Tammie Benzinger, Parinaz Massoumzadeh, Jason Hassenstab, Randall J. Bateman, John C. Morris, Richard J. Perrin, Jasmeer Chhatwal, Mathias Jucker, Bernardino Ghetti, Carlos Cruchaga, Neill R. Graff-Radford, Peter R. Schofield, Hiroshi Mori, and Chengjie Xiong. 2020. "Sequence of Alzheimer Disease Biomarker Changes in Cognitively Normal Adults: A Cross-Sectional Study." *Neurology* 95(23):e3104. doi: 10.1212/WNL.0000000000010747.

Montine, Thomas J., Creighton H. Phelps, Thomas G. Beach, Eileen H. Bigio, Nigel J. Cairns, Dennis W. Dickson, Charles Duyckaerts, Matthew P. Frosch, Eliezer Masliah, Suzanne S. Mirra, Peter T. Nelson, Julie A. Schneider, Dietmar Rudolf Thal, John Q. Trojanowski, Harry V. Vinters, and Bradley T. Hyman. 2012. "National Institute on Aging-Alzheimer's Association Guidelines for the Neuropathologic Assessment of Alzheimer's Disease: A Practical Approach." *Acta Neuropathologica* 123(1):1–11. doi: 10.1007/S00401-011-0910-3.

Porsteinsson, A. P., R. S. Isaacson, Sean Knox, M. N. Sabbagh, and I. Rubino. 2021. "Diagnosis of Early Alzheimer's Disease: Clinical Practice in 2021." *Journal of Prevention of Alzheimer's Disease* 8(3):371–86. doi: 10.14283/JPAD.2021.23/TABLES/5.

Rey A. 1941. "L'examen Psychologique Dans Les Cas d'encéphalopathie Traumatique. (Les Problems.)." *Arch Psychol.* 28:215-285.

Sager, Mark A., Bruce Hermann, and Asenath La Rue. 2005. "Middle-Aged Children of Persons with Alzheimer's Disease: APOE Genotypes and Cognitive Function in the Wisconsin Registry for Alzheimer's Prevention." *Journal of Geriatric Psychiatry and Neurology* 18(4):245–49. doi: 10.1177/0891988705281882.

Segalas, Corentin, Hélène Amieva, and Hélène Jacqmin-Gadda. 2019. "A Hypothesis Testing Procedure for Random Changepoint Mixed Models." *Statistics in Medicine* 38(20):3791–3803. doi: 10.1002/sim.8195.

Sperling, Reisa A., Clifford R. Jack, and Paul S. Aisen. 2011. "Testing the Right Target and Right Drug at the Right Stage." *Science Translational Medicine* 3(111). doi: 10.1126/SCITRANSLMED.3002609.

Team, S. D. 2014. "Stan: A C++ Library for Probability and Sampling, Version 2.0."

Terrera, G. Muniz, A. van den Hout, and F. E. Matthews. 2011. "Random Change Point Models: Investigating Cognitive Decline in the Presence of Missing Data." *Journal of Applied Statistics*

38(4):705–16. doi: 10.1080/02664760903563668.

Terrera, G Muniz, A. Van Den Hout, and F. E. Matthews. 2011. "Random Change Point Models: Investigating Cognitive Decline in the Presence of Missing Data." *Journal of Applied Statistics* 38(4):705–16. doi: 10.1080/02664760903563668.

Thal, Dietmar R., Udo Rüb, Mario Orantes, and Heiko Braak. 2002. "Phases of A Beta-Deposition in the Human Brain and Its Relevance for the Development of AD." *Neurology* 58(12):1791–1800. doi: 10.1212/WNL.58.12.1791.

Thorvaldsson, Valgeir, Stuart W. S. MacDonald, Laura Fratiglioni, Bengt Winblad, Miia Kivipelto, Erika Jonsson Laukka, Ingmar Skoog, Simona Sacuiu, Xinxin Guo, Svante Östling, Anne Ḃrjesson-Hanson, Deborah Gustafson, Boo Johansson, and Lars B̈ckman. 2011. "Onset and Rate of Cognitive Change before Dementia Diagnosis: Findings from Two Swedish Population-Based Longitudinal Studies." *Journal of the International Neuropsychological Society : JINS* 17(1):154–62. doi: 10.1017/S1355617710001372.

Toepper, Max. 2017. "Dissociating Normal Aging from Alzheimer's Disease: A View from Cognitive Neuroscience." *Journal of Alzheimer's Disease* 57(2):331. doi: 10.3233/JAD-161099.

US Department of Health and Human Services. 2018. "Early Alzheimers Disease: Developing Drugs For Treatment, Guidelines for Industry." Retrieved November 29, 2021 (https://scholar-google-com.ezproxy.library.wisc.edu/scholar?q=Early Alzheimers Disease: Developing Drugs For Treatment, Guidelines for Industry).

Wada-Isoe, Kenji, Takashi Kikuchi, Yumi Umeda-Kameyama, Takahiro Mori, Masahiro Akishita, Yu Nakamura, and on behalf of the ABC Dementia Scale Research Group. 2019. "Global Clinical Dementia Rating Score of 0.5 May Not Be an Accurate Criterion to Identify Individuals with Mild Cognitive Impairment." *Journal of Alzheimer's Disease Reports* 3(1):233. doi: 10.3233/ADR-190126.

Wechsler, D. 1987. "WMS-R: Wechsler Memory Scale-Revised." *New York, NY: Psychological Corporation*.

Wechsler, D. 1997. "WAIS-III: Wechsler Adult Intelligence Scale." *New York, NY: Psychological Corporation*.

Williams, Owen A., Yang An, Nicole M. Armstrong, Melissa Kitner-Triolo, Luigi Ferrucci, and Susan M. Resnick. 2020. "Profiles of Cognitive Change in Preclinical and Prodromal Alzheimer's Disease Using Change-Point Analysis." *Journal of Alzheimer's Disease : JAD* 75(4):1169. doi: 10.3233/JAD-191268.

Yang, Lili, and Sujuan Gao. 2013. "Bivariate Random Change Point Models for Longitudinal Outcomes." *Statistics in Medicine* 32(6):1038–53. doi: 10.1002/sim.5557.

Younes, Laurent, Marilyn Albert, Abhay Moghekar, Anja Soldan, Corinne Pettigrew, and Michael I. Miller. 2019. "Identifying Changepoints in Biomarkers during the Preclinical Phase of Alzheimer's Disease." *Frontiers in Aging Neuroscience* 11(APR):74. doi: 10.3389/FNAGI.2019.00074/BIBTEX.

Yu, Lei, Patricia Boyle, Robert S. Wilson, Eisuke Segawa, Sue Leurgans, Philip L. De Jager, and David A. Bennett. 2012. "A Random Change Point Model for Cognitive Decline in Alzheimer's Disease and Mild Cognitive Impairment." *Neuroepidemiology*. 39:73–83.